\documentclass[prb,twocolumn,floatfix,longbibliography]{revtex4-2}

\usepackage{amsmath}  
\usepackage{amsfonts} 
\usepackage{graphicx} 
\usepackage{xcolor}
\usepackage{tikz}

\newif\ifredlined

\redlinedfalse

\ifredlined
\newcommand\circled[1]{\marginpar{\revcolor{red}\tikz[baseline=(char.base)]{\node[shape=circle,draw,inner sep=2pt] (char) {#1};}}}
\newcommand\revcolor[1]{\color{#1}}
\else
\newcommand\circled[1]{}
\newcommand\revcolor[1]{}
\fi

\newcommand{\be}{\begin{equation}}
\newcommand{\ee}{\end{equation}}
\newcommand{\ba}{\begin{align}}
\newcommand{\ea}{\end{align}}

\newcommand\kk{\mathbf{k}}
\newcommand\kt{\mathbf{k}_t}

\begin{document}

\title{Which part of the Brillouin zone contributes most to the high-harmonic radiation?}
\author{M.~Kolesik}

\affiliation{James Wyant College of Optical Sciences, The University of Arizona, Tucson, AZ 85721, U.S.A. }

\begin{abstract}
  Utilizing realistic simulations of high-harmonic generation (HHG) in several materials, we study how different
  regions of the Brillouin zone contribute to the nonlinear response. It is often assumed that
  only the vicinity of the $\Gamma$ point is predominantly responsible for the HHG spectrum, but it is shown here
  that such an approximation is inaccurate in general. While examples can be identified where merely 0.4\% of the
  Brillouin zone produces semi-quantitatively accurate HHG-spectra, in most situations one must include at least thirty to
  fifty percent of the Brillouin-zone volume to obtain accurate above-the-gap harmonics.
  For the harmonic peaks below the bandgap energy, the current-density responses from the entire Brillouin zone must always be integrated.
  We also identify the minimal set of electronic bands necessary for the construction of reduced but still realistic HHG-models.
  The results should be useful for a number of HHG applications, including all-optical reconstructions of the
  band-structure and light-matter couplings, or considerations involving semi-classical approaches to solid-state
  high-harmonic radiation.
\end{abstract}

\maketitle

\section{Introduction}

Ever since the first observations of the above-the-gap harmonic generation from a solid-state medium~\cite{Ghimire11},
numerous efforts continue~\cite{recentHHGtrends} to understand the underlying physics, and to utilize this extremely
nonlinear effect~\cite{ReisReviewHHG,BrabecRev} for applications opening new windows into the
quantum world of materials. The solid-state high-harmonic generation (HHG) has emerged as a tool to map
the band-structures of materials and to investigate their dynamics with an unprecedented resolution.
However, harnessing its full potential as a probe  will require detailed understanding of the
contributions from different  electronic states. In particular, it is necessary to quantify
how different parts of the Brillouin zone contribute to the HHG-signal, and which bands
take part in the dynamics.

The question of ``where in the Brillouin zone is the source'' of the high-harmonic signal, underlines
not only some HHG interpretations but also a number of applications, including all-optical reconstruction of
the electronic band-structure~\cite{ReconstAllOptBands,Lanin:17,SmirnovaIvanov}, characterization of light-matter
couplings~\cite{ReconstDipole,TDP1,TDP2,TDP3,RetrievalD}, and probing the topological properties of
materials~\cite{Baykusheva,Brabec,PhysRevB.106.205422,TunableHuber}. Here, the notion of trajectory,
be it in real or reciprocal space,
often plays a role.

Originally inspired by the success of the strong-field approximation in the HHG from atoms,
the three-step model~\cite{SemiVampa,Vampa_2017,Gaarde22}, its variations~\cite{Tang2022,RecipTraj,Lanin:17} and 
generalizations~\cite{PhysRevLett.127.223201}
were adopted to the solid-state context~\cite{Huang_2022}. A common feature among the strong-field and semi-classical
approaches is that there is an ``excitation step'' when tunneling from the valence to the conduction band occurs at
or close to the $\Gamma$-point (or at the location of the minimal energy gap). 
While it has been pointed out~\cite{PhysRevA.103.063105} that the minimal-gap location does not necessarily dominate
the whole HHG process,  the assumptions that the interband excitation peaks sharply around the point of minimal gap
are rarely tested in concrete situations (however, see~\cite{TunableHuber} for a notable exception).

The importance of electronic trajectories and their
manifestations in the nonlinear response of materials is one of the motivations for this work.
However, beyond the strong-field and semi-classical approaches, quantitative mapping of the Brillouin
zone and of the various contributions to the solid-state HHG is conceptually important in its own right,
and it is our main goal.
This work concentrates on three-dimensional materials, and utilizes realistic 
numerical simulations to compare how different regions of the Brillouin zone shape the
high-harmonic responses.

Our results will show that there is a qualitative difference between the behaviors of the
below-gap and the above-gap harmonics. While for the lower-order harmonic radiation the contributions
from the entire Brillouin zone are always necessary to achieve quantitatively accurate description,
only thirty to fifty percent of the Brillouin zone volume is sufficient to capture a great majority of the
higher-order harmonic radiation. However, it turns out that, at least for the three-dimensional
materials, the assumption that the electronic states from around the $\Gamma$ point contribute
the most of the HHG response is unrealistic in general.

Such a finding may seem to contradict the fact that the tunneling is by far most likely where the
energy gap is minimal. To clarify this issue, we investigate what portion of the excited carriers
are actually ``born'' in the central part of the Brillouin zone. It will be shown that despite the local
tunneling rate being much larger around $\Gamma$, the rest of the zone still generates the great majority
of the excited carriers. Their collective response to the driving field therefore cannot be always neglected.

We  also address the question concerning the minimal set of electronic bands that should be included
in a reduced model. While some spectral features can be obtained already
with a few bands, accurate results require inclusion of bands in sets that are ``connected'' via
degenerate points or via avoided band crossings.

\section{Material models}

This work utilizes the empirical tight-binding models for several materials with zinc blende and diamond structures.
Most of our simulations were performed for GaAs. Besides the fact that it is an important material,
this choice is also motivated by the fact that many experimental results are available for this material
(see e.g.~\cite{PhysRevB.104.L121202,Xia:18}).
Some of the measurements~\cite{Xia:18} were previously used to verify the accuracy of the model used in this
paper. We have demonstrated that the tight-binding description of GaAs, both with and without spin-orbit
coupling, provides the nonlinear optical response which agrees quite well with HHG measurements~\cite{KolesikPRB}.
It was also shown that the model correctly captures the second-order nonlinearity of  the material~\cite{Kolesik:23}.
In other words, the tight-binding description of GaAs, coupled with the recently introduced HHG simulator~\cite{AlgSBE}
has been tested across the frequency range including below-bandgap and above-bandgap harmonics.
This gives us confidence to draw conclusions from the numerical simulations.

The next material included in our comparative simulations is CdTe. It was recently identified as
a medium in which harmonics up to and beyond order of thirty were generated at a very
low excitation intensity~\cite{ExpCdTe}. The low-intensity and flux needed for the high-order harmonic
generation makes this material very interesting. One reason this medium is suitable for the present purpose
is that it complements GaAs. It was argued in Ref.~\cite{ExpCdTe} that the special feature that
makes CdTe to stand apart is the flatness of its conduction band. The flatness implies a  high density of
states in the vicinity of the $\Gamma$-point, which in turn makes it possible to argue that the center
of the Brillouin zone should command the HHG process even more than in GaAs.
The strength of the $\Gamma$-point contribution is precisely the feature which we aim to investigate.

To broaden our set of material models, we choose crystalline Silicon as a material which differs from
GaAs and CdTe in that is is inversion symmetric. The higher material symmetry eliminates even-order
harmonic generation, and gives rise to simpler spectra in which it is perhaps easier to investigate
which part of the Brillouin zone contributes most.
Unlike GaAs, crystalline Silicon does not possess a direct gap. Consequently, it is not straightforward
to make an argument for a specific location in the Brillouin zone to be the strongest ``source'' of
high-harmonic radiation. It should be interesting to see if direct- and indirect-gap media
behave differently in this respect.

The description of the materials in this work is based on the same, so called sp$^3$s$^*$ tight-binding model~\cite{VOGL}.
It is applicable to both the zinc blende and the diamond structures. As an input for the sgiSBE
solver~\cite{AlgSBE,Kolesik:23}
which was used to simulate the HHG spectra, the explicit expressions for the $\mathbf{k}$-dependent Hamiltonian $h(\mathbf{k})$
was obtained from~\cite{TBM1}. The parameter sets were taken from~\cite{VOGL} for GaAs and Silicon,
and from~\cite{CdTeTBM} for CdTe. For the results shown in what follows, we applied the spin-degenerate version of the
material models.

\section{Numerical modeling of solid-state HHG}

HHG simulations in this work were done with the structure-gauge-independent SBE solver (sgiSBEs).
We refer the Reader to the descriptions of the algorithm given in~\cite{AlgSBE} and also in \cite{Kolesik:23},
and include a brief summary here.
For each $\kk$ sampling the Brillouin zone, the initial density matrix $\rho_{mn}( \mathbf{k} ; t=0)$
is set to be the zero-temperature density matrix representing full valence and empty conduction bands. 
One integration step, or an update from time $t_i$ to time $t_{i+1}$, can be written as
an operator splitting scheme in the form 
\begin{align}
\label{eq:algo}
\rho_{mn}( \mathbf{k} ; t_{i+1}) = \sum_{a,b}
&\langle m \mathbf{k}_{i+1}\vert  a \mathbf{k}_i\rangle \times \nonumber \\
&e^{-i \epsilon_a(\kk_i) \Delta }   \rho_{ab}(\kk ;t_{i}) e^{+i \epsilon_b(\kk_i) \Delta } \times \nonumber \\
&\langle b \mathbf{k}_i   \vert n \mathbf{k}_{i+1}\rangle  \ .
\end{align}
where $\Delta=t_{i+1} - t_i$ is the time-step,  $\mathbf{k}_i = \kk - \mathbf{A}(t_i)$ with
$\mathbf{A}$ representing the vector potential of the driving pulse, and
$\vert  a \mathbf{k}_i\rangle$ stands for the $\kk$-dependent Bloch eigenstate in
the electronic band $a$. The middle row in (\ref{eq:algo}) can be understood as
the first split-step which evolves the density matrix in the current Hamiltonian
eigenbasis $\{\vert  a \mathbf{k_i}\rangle\}$. For the next split-step,
the new-time eigenbasis  $\{\vert  b \mathbf{k}_{i+1}\rangle\}$ is obtained by the
exact diagonalization of the Hamiltonian $h(\mathbf{k}_{i+1})$, and then the density matrix
is transformed into the new basis as shown in the first and third rows of (\ref{eq:algo}).
This update scheme is applied in parallel, independently for all $\kk$ sampling the Brillouin zone.
Dephasing is approximated as usual, in a separate split-step,
\be
\rho_{mn}(\kk ; t) \leftarrow \rho_{mn}(\kk ; t)  \exp\left[-\Delta t/T_2 \right] \ ,
\ee
with $T_2$ being the dephasing time, which we set equal to 5~fs, and note that the observations
we arrive at in this work are independent of its precise value.

 The total observed current density is calculated as an
integral over the complete Brillouin zone, 
\be
\mathbf{j}(t) =
\sum_{mn}\int\displaylimits_\text{BZ}\frac{d\kk}{(2\pi)^3} \langle n\kt\vert \partial_{\kt} h(\kt)\vert m \kt\rangle \rho_{mn}(\kk ;t) \ ,
\label{eq:observeJ}
\ee
where the Hamiltonian gradient $\partial_{\kt} h(\kt)$  is obtained from the  tight-binding model in an analytic
form, and its matrix element in the above formula represents the operator observable corresponding to the total
current density (for simplicity, we do not decompose $\mathbf{j}(t)$ into intra- and inter-band currents).

An important aspect of the method is the grid of $\kk$-vectors sampling the Brillouin zone,
which must be sufficiently dense for the integral (\ref{eq:observeJ}) to converge.
The convergence properties with respect to the Brillouin-zone
sampling density were discussed in detail in~\cite{KolesikPRB}. For this work, we sample
the reciprocal-space cell with an equidistant grid of 128$^3$ grid points. The grid
is aligned with the reciprocal basis vectors and it is invariant under the symmetry operations
of the material.

\section{Simulation results}

\subsection{Response from a portion of the Brillouin zone}

For a crystalline sample with a fixed orientation, we assume a linearly polarized
optical pulse driving the current-density response, which is subsequently
converted into a spectrum.
For a given $\kk$, the integrand in (\ref{eq:observeJ}) represents the contribution of this $\kk$-spot inside the
Brillouin zone to the total response. Because it is often assumed that  $\kk$ from a close vicinity of $\Gamma$
contribute most to the HHG spectrum, we define a partially-sampled current density as
\be
\mathbf{j}(r,t) =
\sum\!\!\!\!\!\!\int\displaylimits_{\text{BZ}, |\kk|< r}\!\!\!\!\!\!\!\frac{d\kk}{(2\pi)^3} \langle n\kt\vert \partial_{\kt} h(\kt)\vert m \kt\rangle \rho_{mn}(\kk ;t) \ .
\label{eq:partJ}
\ee
In other words, for a given ``radius'' $r$ we only include the current-density contributions from the
initial $\kk$-vectors that are closer to $\Gamma$ than $r$.
The central question in what follows is how large $r$ one must take for $\mathbf{j}(r,t)$ to be an accurate approximation
of the total current density  $\mathbf{j}(t)$. 

\subsection{HHG from a vicinity of the $\Gamma$-point}

Motivated by the  assumption that high-harmonic radiation is
mostly mediated by the electronic states close to the $\Gamma$-point
of the Brillouin zone, we first look at the spectra generated by such states.
We select a portion of the zone with a radius of $r=1$ (in units of inverse
lattice constant), and calculate their collective contribution to $\mathbf{j}(r=1,t)$
and then the corresponding  HHG-spectrum.
During this and the following simulations, we make sure that all symmetry-related
$\mathbf{k}$-points are simultaneously included, so that the calculated
response has the correct material symmetry. This particular (i.e. for $r=1$) symmetry-respecting
``$\Gamma$-vicinity'' contains
only about 0.4\% of the total Brillouin-zone volume, so it is rather small.

As a first example, take GaAs excited by a 3.5$\mu$m pulse with the electric-field
amplitude of 8.7MV/cm, polarized along the [011] direction. For this sample orientation,
one expects the odd harmonics to be $p$-polarized, i.e. parallel to the excitation,
while the even harmonics should be $s$-polarized.

\begin{figure}[t]
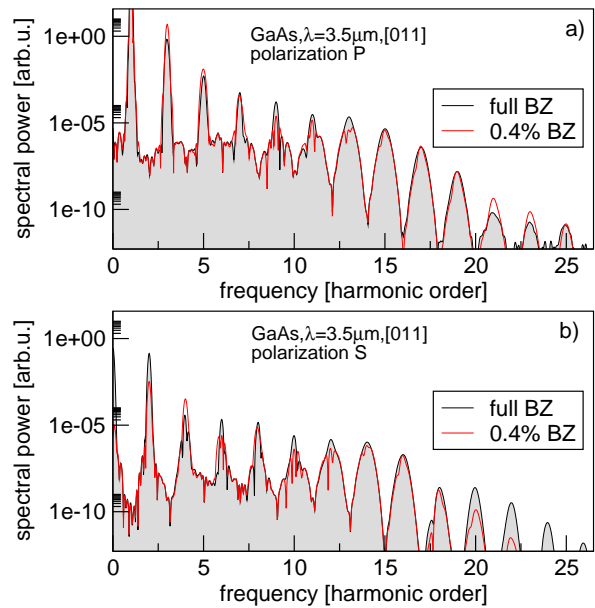

\centerline{\includegraphics[clip,width=0.9\linewidth]{./figure1a.eps}}
\centerline{\includegraphics[clip,width=0.9\linewidth]{./figure1b.eps}}
\caption{
  GaAs high-harmonic spectrum from the full Brillouin zone (black, gray-filled line)
  compared to that generated from a vicinity of $\Gamma$-point encompassing 0.4\%
  of the total Brillouin-zone volume (red line), for a) parallel and b) perpendicular
  polarization. The excitation pulse, at $\lambda=3.5\mu$m wavelength, is polarized
  along $y=z$ direction.
\label{fig:1}
  }
\end{figure}

Figure~\ref{fig:1} shows the simulated HHG-spectra, generated by the electronic states
that are initially in the $\Gamma$-vicinity, compared to the spectrum
in which the entirety of the Brillouin zone was included. For the $p$-polarization,
the two HHG-spectra are quite similar. Keeping in mind that the scale of the
figure is logarithmic, one can see that the largest deviations between the two
spectra are about one order of magnitude or only slightly larger. For the $s$-polarized
response, the gap between the partial and total spectra opens a bit more, especially
for the second harmonic and the harmonics above the order of eighteen. Nevertheless,
the fact that merely 0.4\% of the Brillouin zone volume produces a HHG response
so close to the total suggest that the $\Gamma$-point indeed dominates
the HHG process as is often assumed in various semi-classical analyses.

\begin{figure}[t]
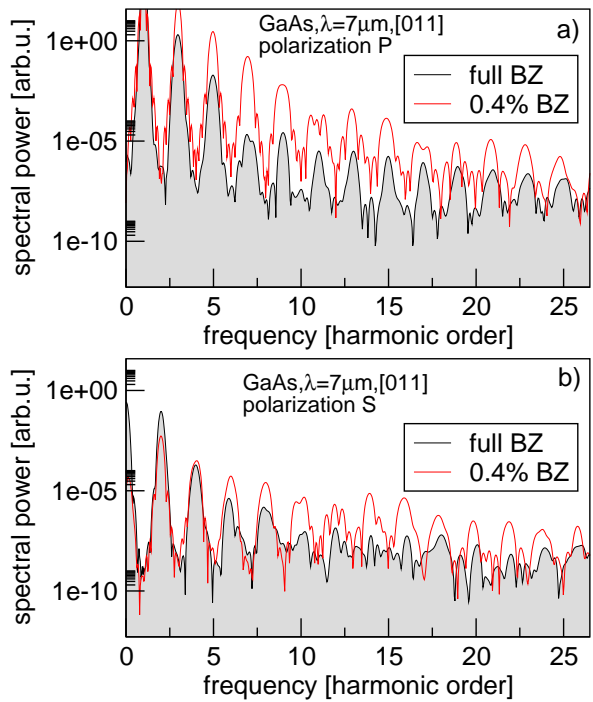

\centerline{\includegraphics[clip,width=0.9\linewidth]{./figure2a.eps}}
\centerline{\includegraphics[clip,width=0.9\linewidth]{./figure2b.eps}}
\caption{
  GaAs high-harmonic spectrum from the full Brillouin zone (black, gray-filled line)
  compared to that generated from a vicinity of $\Gamma$-point encompassing 0.4\%
  of the total Brillouin-zone volume (red line), for A) parallel and B) perpendicular
  polarization. The excitation pulse, at $\lambda=7\mu$m wavelength, is polarized
  along $y=z$ direction.
\label{fig:2}
  }
\end{figure}

This is an encouraging observation, but it would be premature to conclude that
the prominence of the $\Gamma$-point as the HHG-source is a typical behavior.
For the next example, the excitation wavelength is chosen twice as long, $\lambda=7\mu$m.
The expectation is, perhaps, that in response to longer wavelengths a small vicinity of the
Brillouin-zone center also generates most of the harmonic power; Indeed, one could
argue that a longer wavelength implies a more off-resonant tunneling excitation
of electrons into the conduction bands, which should translate into carriers being
``born'' in a tighter neighborhood of $\Gamma$. As a result, even smaller portion
of the Brillouin zone could be responsible for the harmonic response.

However, Fig.~\ref{fig:2}
shows otherwise. While all conditions with the exception of the excitation wavelength
are exactly the same as in the former example, one can see a dramatic difference
between the partial and full spectra. Perhaps one should consider a larger volume around
the center of the zone, but in the light if the previous illustration, this result is
rather surprising. Yet, it turns out that it is more typical...

\begin{figure}[t]
  \centerline{\includegraphics[clip,width=0.9\linewidth]{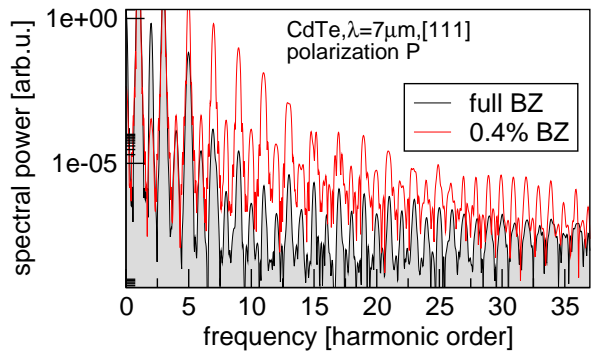}}
\caption{
  CdTe high-harmonic spectrum from the full Brillouin zone (black, gray-filled line)
  compared to that generated from a vicinity of $\Gamma$-point encompassing 0.4\%
  of the total Brillouin-zone volume (red line).
  The excitation pulse, at $\lambda=7\mu$m wavelength, is polarized
  along $x=y=z$ direction.
\label{fig:3}
  }
\end{figure}

Figure~\ref{fig:3} shows an analogous comparison of partial and full-zone
spectra for CdTe. A  recent study showed that the high-harmonic generation
is significantly stronger in this material, and the relatively flat conduction-band
shape was identified as a possible reason. The higher density of states around $\Gamma$
should further emphasize its contribution to the response, and this is why the material
was selected for our next example. Here we have chosen to change the excitation-pulse
polarization to [111], for which both even and odd
harmonics appear in the $p$-polarization, while the $s$-polarized response vanishes.
The figure shows a range of frequencies similar to that observed
in the experiments~\cite{ExpCdTe}. In a qualitative agreement with the experiment, a rather strong
set of harmonics forms between order 20 and 30, and their relative powers appear to
be in the right ballpark. We therefore trust that this simulation, too, is sufficiently
realistic. The important takeaway from this numerical experiment is that the small central portion
of the Brillouin zone provides a poor approximation of the actual HHG spectrum.
While the qualitative ``shape'' and relative peak powers are qualitatively
very similar between the two spectra, the partial response is orders of magnitude
{\em stronger}.

\begin{figure}[t]
  \centerline{\includegraphics[clip,width=0.9\linewidth]{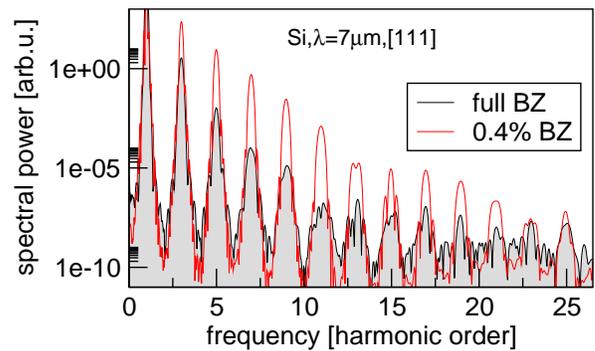}}
\caption{
  Silicon high-harmonic spectrum from the full Brillouin zone (black, gray-filled line)
  compared to that generated from a vicinity of $\Gamma$-point encompassing 0.4\%
  of the total Brillouin-zone volume (red line).
  The excitation pulse, at $\lambda=7\mu$m wavelength, is polarized
  along $x=y=z$ direction.
\label{fig:4}
  }
\end{figure}

As yet another example, we choose crystalline Silicon as an inversion-symmetric material
with an indirect band-gap. The sample orientation and the properties of the excitation pulse
are the same as in the previous case, i.e. $\lambda=7\mu$m and the electric field oscillation
direction is [1,1,1].
Figure~\ref{fig:4} shows, once again, that the immediate neighborhood of $\Gamma$ does not
produce a HHG response which could be compared to that from the full Brillouin zone.
One could even argue that the difference here is larger than that in the previous examples.
Indeed, the harmonic orders 7 --- 11 appear to be four to five orders of magnitude
stronger than those generated from the full Brillouin zone. We speculate that this
could be the manifestation of the indirect gap of this material.

To summarize this subsection, we have shown that the high-harmonic spectra generated from
a small neighborhood of the Brillouin zone center are not, at least not in general, good
approximation of the full response. Thus, one needs to go beyond the immediate vicinity
of the $\Gamma$-point to obtain a more precise representation of the HHG-spectra.
In the following subsection we quantify
how large the integration region should be.

\begin{figure}[t!]
  \centerline{\includegraphics[clip,width=0.9\linewidth]{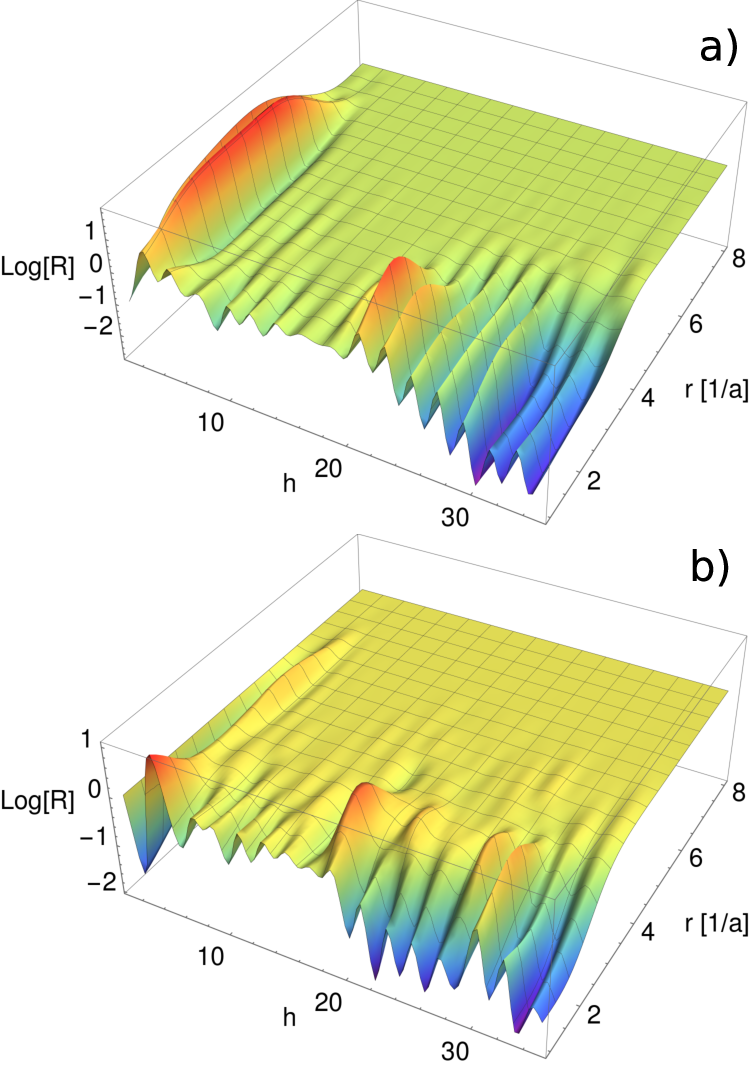}}
\caption{
  Convergence  of the simulated spectrum for GaAs excited along [011] at $\lambda=3.5\mu$m
  (as in Fig.~\ref{fig:1}). Quantity  $\log[R(r,h)]$ (see text for details),
  representing the ratio of the partial ($r<7.025...$) to full-zone spectral power
  integrated over the harmonic band $h$  is shown on a log scale.
  Top (a) and bottom (b) panels represent the results for the $p$- and $s$-polarized HHG spectra,
  respectively.
\label{fig:5}
  }
\end{figure}

\subsection{Mapping the HHG-sources}

In order to map the strength of the HHG source across the Brillouin zone, we
set up a series of simulations for each of the studied materials.
We increase the radius $r$ of the $\Gamma$-centered subset of the Brillouin
zone in steps of 1$a^{-1}$, calculating the current-density $\mathbf{j}(r,t)$ and the corresponding
spectrum $S(r,\omega)$ at each step, until
the entirety of the zone is included. Note that the values
$r=\{1,2,3,4,5,6,7\}a^{-1}$ represent, respectively,
0.4, 3.4, 11, 27, 54, 86, and 99.99 percent of the full-zone volume.
The goal of this mapping is to identify how large a
portion of the Brillouin zone must be included for a reasonably accurate
HHG simulation.

\begin{figure}[t]
  \centerline{\includegraphics[clip,width=0.9\linewidth]{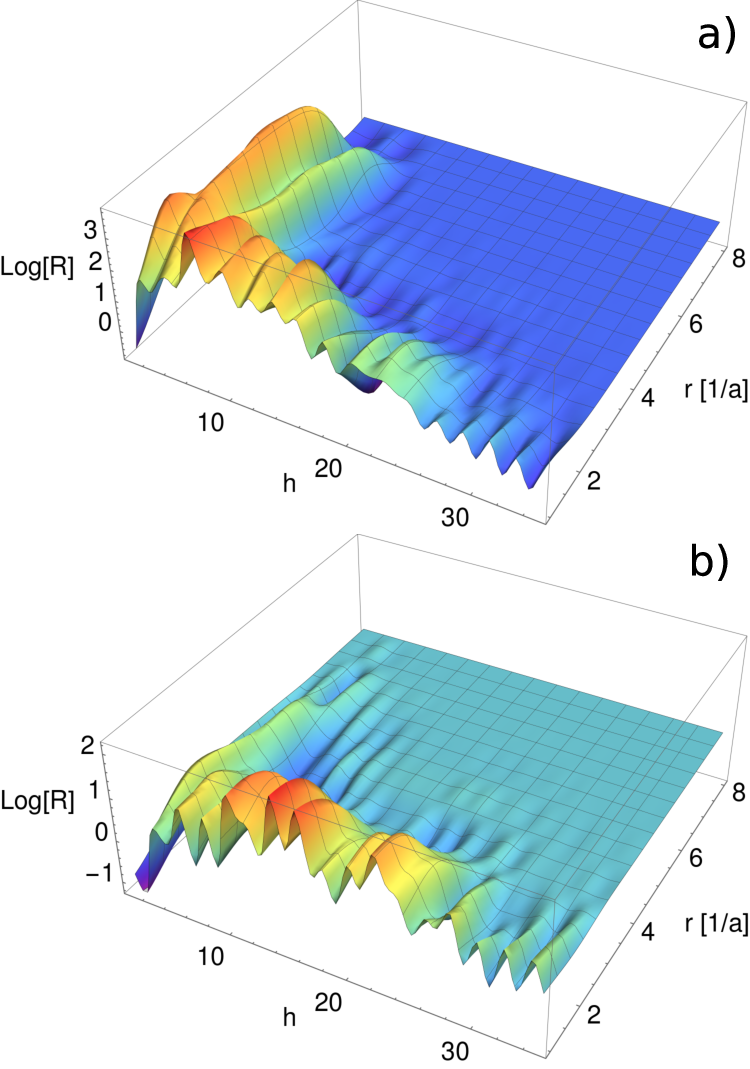}}
\caption{
 Simulated spectrum convergence for GaAs excited along [011] at $\lambda=7\mu$m
  (as in Fig.~\ref{fig:2}). a) $p$-polarized response, and a) $s$-polarized response.
\label{fig:6}
  }
\end{figure}

For the sake of visualization, the integrated harmonic-band power $P(r,h)$
is calculated from the simulated spectrum $S(r,\omega)$
in the given simulation as
$P(r,h)=\int_{(h-0.5)\omega_0}^{(h+0.5)\omega_0} \vert S(r,\omega)\vert^2 d\omega$. The integration
interval is always centered on the integer multiple of the fundamental frequency $\omega_0$,
whether that particular harmonics is allowed by symmetry or not, and the interval is one harmonic order
wide. The two-dimensional maps of $R(r,h) = P(r,h)/P(r=8,h)$ are plotted versus
the order $h$ and $r\in (1,8) a^{-1}$. This quantity or, more precisely, its approach to unity, reflects the spectrum
convergence since for $r=8$ the entire Brillouin zone is already included
in the integration. (Note that $r\approx 7.025$, corresponding to $\vert\Gamma W\vert$,
represents the entire zone.) To make the resulting graphs easier to interpret,
at least for their global features, the interpolation order of the plot routine
was set equal to three (resulting in an artificially smooth plotted surface). Since the log-scale
is natural for simulated spectra, $\log[R(r,h)]$ is depicted in the following two figures.

Figure~\ref{fig:5} illustrates the convergence of the simulated spectrum with the
increasing radius $r$ of the portion of the Brillouin zone included in the simulation.
The material here is GaAs and all conditions as for Fig.~\ref{fig:1}. Thus, this figure shows
how the red curves in Fig.~\ref{fig:1} gradually approach (with the increasing $r$) the
black curves corresponding to the full-zone spectrum.

Several features become evident; First, the below-gap harmonics are never accurate
until $r\approx 7 a^{-1}$, i.e. essentially until the entire zone is included in the
simulation. Second, for the higher-order harmonics it takes $r\approx 4 a^{-1}$, i.e.
about 25\% of the full-zone volume to obtain accurate simulated spectrum.
Third, while quantitatively different, the convergence behaviors of the $p$- and $s$-polarized responses
are similar.

Figure~\ref{fig:6} shows similar results for the longer excitation wavelength,
$\lambda=7\mu$m, and corresponds to the transition between the extreme cases depicted
in Fig.~\ref{fig:2}. Comparison to Fig.~\ref{fig:5} shows that while the
small-$r$ results are an order of magnitude less accurate (note the different
vertical scale in the two figures), a good convergence (but for higher-order
harmonic only) is achieved at about the same rate, and perhaps even faster
at the high-frequency end of the spectrum.
We note that the convergence behavior is very similar also in the cases
of Silicon and CdTe, illustrated in Figures~\ref{fig:3} and \ref{fig:4},
and therefore their convergence-maps are not shown here.

\begin{figure}[t]
    \centerline{\includegraphics[clip,width=0.9\linewidth]{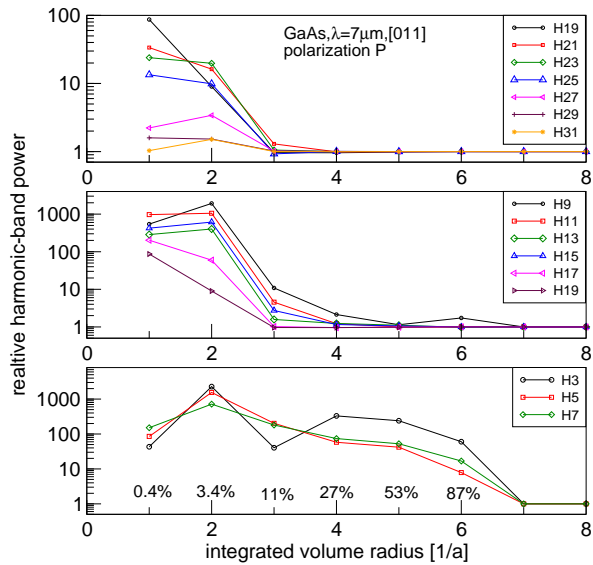}}
\caption{
  Convergence of the $p$-polarized HHG spectrum for GaAs excited by a pulse polarized along $y=z$
  at $\lambda=7\mu$m. 
  Shown is the ratio $R(r,h)$ between the spectral power (integrated over
  a harmonic band) obtained from a portion of the Brillouin zone and the
  corresponding power for the entire zone. Lines connecting the symbols
  only serve as a guide for the eye. Percentages shown in the lower panel
  represent the volume-portion of the whole zone. 
\label{fig:7}
  }
\end{figure}

\begin{figure}[t!]
  \centerline{\includegraphics[clip,width=0.9\linewidth]{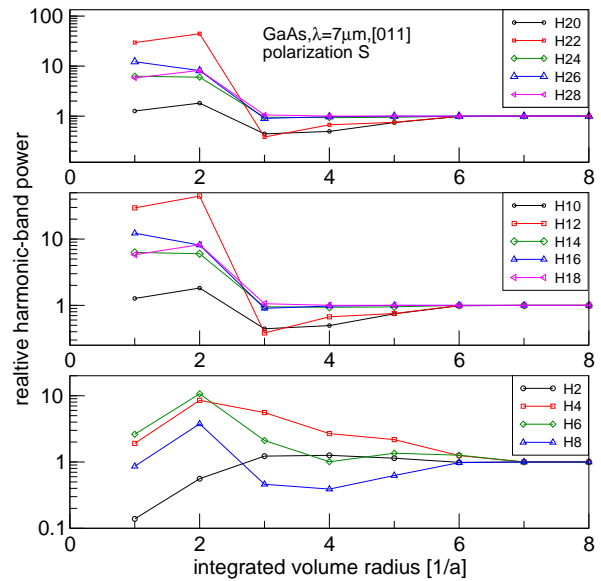}}
\caption{
  Convergence of the $s$-polarized HHG spectrum for the same conditions as
  in Fig.~\ref{fig:7}.
\label{fig:8}
  }
\end{figure}

For a more quantitative representation of the convergence of the simulated spectra
as functions of the portion of the Brillouin zone included, Figs.~\ref{fig:7} and~\ref{fig:8}
show the ratio $R(r,h) = P(r,h)/P(r=8,h)$ between the partial-spectrum and full-spectrum power integrated over the
given harmonic band. The bottom panels show that the lower-order harmonics can not be obtained
accurately unless the entire zone is integrated. The medium-high harmonic orders
(shown in the middle panels) converge
faster, with the partial-zone results accurate within an order of magnitude once the
integrated zone-portion is larger than about one third. The high-order harmonics (top panels)
become quite accurate already at about 25 to 50 percent of the entire zone included.

So, to answer the question about the size of the zone that needs to be included in the
simulation, it is hundred and roughly fifty percent for the below-gap and the above-gap harmonics,
respectively. From the simulation standpoint, a fifty-percent increase in speed is
probably not worth of the trouble...

On the other hand, this result is important for the interpretations of HHG dynamics in those
cases where any kind of trajectory comes into consideration. It indicates that the starting
points of the contributing trajectories should ``cover'' at least thirty to fifty percent of the
Brillouin-zone volume, and that the corresponding contributions should be coherently added up
to represent the total response of the material. 

\subsection{Mapping the tunneling rate}

\begin{figure}[t!]
  \centerline{\includegraphics[clip,width=0.9\linewidth]{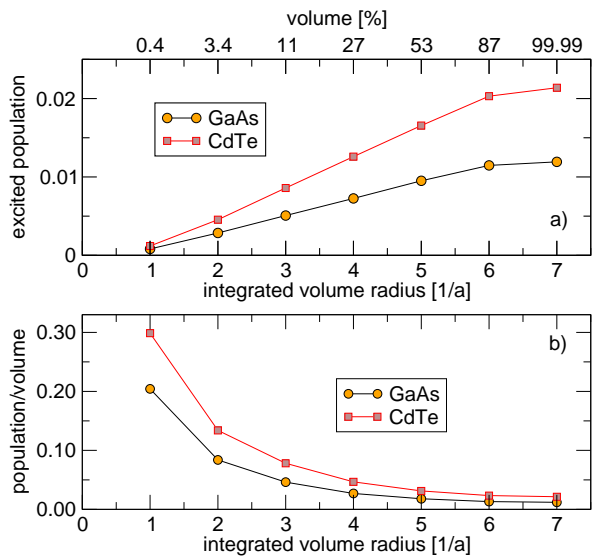}}
\caption{
  Total population fraction of excited states inside the reciprocal-space integration volume
  centered on the $\Gamma$-point (a). The top horizontal axis shows the size of the volume as
  a percent-fraction of the full Brillouin zone. The lower panel, (b) shows the population divided
  by the volume which gives the average tunneling rate for the states inside the volume of radius $r$. As expected,
  the tunneling rate peaks in the center of the zone.
\label{fig:9}
  }
\end{figure}

The finding that the Brillouin zone center in general does not contribute the most of the HHG
response seems to contradict the highest tunneling probability in this region. Indeed, if the
valence-to-conduction band excitation does behave as the tunneling ionization in atoms, then the
rate of excitation should be exponentially sensitive to the k-dependent bandgap, and therefore fall-off
quickly upon departure from the zone center. Based on this
argument, one sometimes assumes that the carriers are only created {\em at} the point of the minimal gap.

The explanation is actually very simple. It turns out that the difference between the tunneling rates
is counteracted by the fact that the volume of the Brillouin zone overshadows the vicinity of the
$\Gamma$ point. To illustrate this, we have measured the total population in all conduction bands combined.
Figure~\ref{fig:9} shows the result for the examples of GaAs and CdTe excited with a $\lambda =7\mu$m pulse.
The top panel makes a case for the great majority of carriers being generated away from the $\Gamma$ point.
The lower panel depicts the total excited population in the integrated subset of the Brillouin zone
divided but the volume of latter. As such it is a measure of the (averaged) tunneling rate. This quantity
indeed shows that the excitation rate is greatly larger around the zone-center, precisely as expected.
It is just that the contrast is not large enough to compensate for the much bigger volume of the
peripheral regions of the Brillouin zone.

\subsection{The relevant energy bands}

One sometimes makes an argument that a HHG spectrum, or perhaps just one of its peaks is
generated by the contributions from a limited number of electronic bands, possibly from only
a single valence-conduction pair. Indeed, this kind of an assumption underlines a number of
approaches for all-optical reconstruction applied to the band-structures or to the
k-dependent dipole moments.

\begin{figure}[t!]
  \centerline{\includegraphics[clip,width=0.9\linewidth]{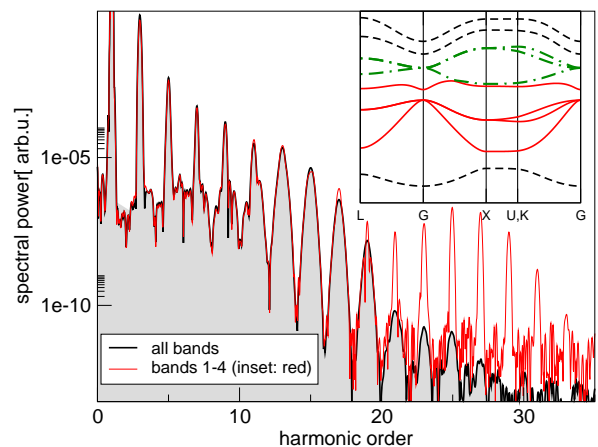}}
\caption{
  High-harmonic spectrum from a reduced model. The black-line, shaded-area shows
  the spectrum obtained from the full simulation. The thin red line represents
  the result of a simulation restricted to the bands shown in full red lines in the inset.
\label{fig:10}
  }
\end{figure}

To test which electronic bands give relevant contributions
in the materials studied here, we have repeated the simulation with restricted sets
of states. The restriction is realized by projecting the density matrix
(or equivalently, by projecting the k-dependent Hamiltonian $h(\mathbf{k})$) onto
a subspace spanned by the select bands at each integration step.

The results will be illustrated on the case of GaAs, excited with a $\lambda = 3.5\mu$m pulse
(the situation corresponding to that illustrated in Fig.~\ref{fig:1}), but the observations
are similar in all situations investigated in this work.

First, we
select the bands 1 through 4 (labeling the lowest-energy band as 0). This selection accounts for the
three highest valence bands and the lowest-energy conduction band. It seems reasonable to expect
that these are the most important bands above and below the gap. We compare the
resulting HHG spectrum to that obtained from the full simulation in Fig.~\ref{fig:10}
for the $p$-polarized component of the response.
It is encouraging that the spectra are quite close to each other for harmonics lower than nineteen. However,
one can see pronounced artifacts at the high-frequency end of the spectrum, where strong harmonic
peaks appear which do not exist or which are
about five orders of magnitude weaker in the full spectrum.

So if we ask which bands need to be included such that the corresponding spectrum is
accurate across all frequencies, the set of one to four appears to be inadequate.
We speculate that the reason is that the conduction band four makes a close approach to the group
of higher-energy bands shown by dashed-dotted line in the inset of Fig.~\ref{fig:10}.
In a time-dependent external field they become coupled and the populations and polarizations resulting from this coupling
affect the dynamics.

Inclusion of the next conduction band does not improve the situation (data not shown);
the artifacts at the higher-frequency side of the spectrum actually become slightly worse.
Only after including the whole group five to seven (dark green dashed-dotted lines
in Fig.~\ref{fig:10}) we obtain a HHG spectrum that is very close to the full spectrum
with the exception of small deviations for the lowest-order harmonics.

So we come to the conclusion that while an individual harmonic peak could be approximated
with a limited set of electronic bands, to guarantee the accuracy for all harmonics,
one should include all ``energetically connected'' bands below and above the bandgap.

\section{Conclusion}

Numerical simulations of the high-harmonic generation in zinc blende materials
GaAs and CdTe, and in crystalline Silicon were executed for two different excitation wavelengths
while gradually including the nonlinear-response contributions from larger
and larger portions of the Brillouin zone. The goal of this work was to determine 
how large is the region in the Brillouin zone that must be accounted for
in a numerical simulation in order to produce quantitatively accurate spectra.
It turns out that there is no
simple answer to the question posed in the title of this work. Nevertheless,
our comparative simulations offer useful insights into certain trends and behaviors
which seem quite universal.

The assumption that the Brillouin zone center dominates the high-harmonic processes
has been used in a number of HHG-interpretations, especially those incorporating aspects
of the semi-classical approaches. However, we have found that this assumption is
not justified, at least not in general.
While we have identified an example where the $\Gamma$-neighborhood encompassing
merely 0.4 percent of the Brillouin zone volume does provide a HHG spectrum
accurate within one order of magnitude globally and with a much better accuracy
locally (specifically for the medium-order harmonic orders), we have also found several cases
where the contribution from the zone center is orders of magnitude different
from the total spectrum. Interestingly, in such cases the central contribution
appears to be {\em stronger} than that of the whole zone, and this indicates that
destructive interferences occur across the Brillouin zone.

On the other hand, it turns out that for (only) the above-the-gap frequencies, the center of the zone
produced {\em qualitatively correct} spectra in all cases studied in this work. The
zone-center HHG spectra exhibited stronger peak-to-valley contrast, but the overall
spectrum shapes were good. We therefore think that, in the light of these results, the
approximations restricted to the  $\Gamma$ point (or to a small vicinity of it) remain justified
for {\em qualitative} investigations into underlying physics. However, it should  be emphasized
that no matter how small a portion of the Brillouin zone is used, the sampling grid must
have the symmetry of the material.

For the simulations which aim for a higher accuracy, we conclude that a significant portion
of the Brillouin zone must be integrated over. Of course, precisely how large it should be
depends on the expected accuracy, material, and the excitation wavelength. As a rough estimate,
thirty to fifty percent of the zone volume should be included for a simulated spectrum with the harmonic peaks accurate
within an order of magnitude or better.
In terms of the excitation wavelength, we observed that while the spectra generated from the zone
center were more accurate for the shorter wavelength, the convergence toward the full-zone
solution seems somewhat faster for the longer wavelength, i.e. a smaller part of the Brillouin zone can
suffice. We speculate that this is because the carriers are driven further across the
Brillouin zone in the latter case.

We trust that these observations will prove useful for a number of applications
concerning solid-state high-harmonic generation, including all-optical band-structure
reconstruction and/or optical measurements of Berry curvatures and shift vectors.
Analyses which rely on the assumption of specific starting points for the relevant trajectories,
could be generalized to account for the fact that in reality a ``distribution of,'' or a bundle
of trajectories should provide a physically more realistic picture.

Another finding with potential impact on such applications is that care should be exercised when
one wants to restrict the number of electronic bands believed to account for the relevant
response contributions. We have found that in order to obtain good accuracy across all frequencies,
all bands that are connected or that closely approach each other somewhere in the Brillouin zone must be
included. In other words, it should be safer to apply various reconstruction algorithms to the
whole ``connected groups'' of electronic bands.

A rather surprising outcome of this study is that the lower-order, below-the-gap harmonics
always require the integration of all response-contributions across the entirety of
the Brillouin zone. This finding is important  for the future microscopic-level
modeling of laser materials, such as zinc-blende wide gap semiconductors. Whenever
the nonlinear propagation of optical pulses plays a role, the second- and third-harmonic responses
in particular must be sufficiently accurate. This work shows that the Brillouin zone center alone
is insufficient to yield accurate description for the nonlinear properties at frequencies below the
gap.

Our work concentrated on the three-dimensional materials, and it is not a given that
the results apply to the (effectively) two-dimensional systems, where the central
part of the Brillouin zone constitutes a much larger fraction of its volume. On one hand,
one could argue that such an investigation is less important form the practical standpoint,
because even complete sampling of two-dimensional Brillouin zones does not present any significant
numerical challenge. On the other hand, it will be interesting to extend the present
study to two-dimensional materials for the conceptual reasons, for example to
improve the semi-classical interpretations.

\section*{Acknowledgments}

Authors acknowledge the support from the
Air Force Office for Scientific Research under grants no.
FA9550-22-1-0182
and
FA9550-21-1-0463.


%

\end{document}